\documentclass[12pt, draftclsnofoot, onecolumn]{IEEEtran} 

\usepackage{newtxtext}
\usepackage{newtxmath}

\usepackage{textcomp}

\usepackage{graphicx}
\usepackage[caption=false,font=footnotesize]{subfig}  

\usepackage{booktabs}
\usepackage{multirow}
\usepackage{tabularx}
\usepackage{makecell}
\usepackage{array}

\usepackage{algorithm}
\usepackage{algpseudocode}

\usepackage{url}
\usepackage{enumitem}
\usepackage{xcolor}
\usepackage{listings}
\usepackage{siunitx}
\usepackage[english]{babel}    

\usepackage{cite}              

\begin{document}


\title{Enhancing Satellite Quantum Key Distribution with Dual-Band Reconfigurable
Intelligent Surfaces}
\author{
Muhammad Khalil,~\IEEEmembership{Member,~IEEE}, 
Ke Wang,~\IEEEmembership{Senior Member,~IEEE},\\ 
and Jinho Choi,~\IEEEmembership{Fellow,~IEEE}%
\thanks{
M. Khalil and K. Wang are with the School of Engineering, RMIT University, Melbourne, VIC 3000, Australia (e-mail: muhammad.khalil@rmit.edu.au; ke.wang@rmit.edu.au).\\
J. Choi is with the School of Electrical and Mechanical Engineering, University of Adelaide, Adelaide, SA 5005, Australia (e-mail: jinho.choi@adelaide.edu.au).}
}
\maketitle
\begin{abstract}
This paper presents a novel system architecture for hybrid satellite
communications, integrating quantum key distribution (QKD) and classical
radio frequency (RF) data transmission using a dual-band reconfigurable
intelligent surface (RIS). The motivation is to address the growing
need for global, secure, and reliable communications by leveraging the
security of quantum optical links and the robustness of classical
RF channels within a unified framework. By employing a frequency-selective
RIS, the system independently optimizes both quantum (850 nm) and
classical (S-band) channels in real time, dynamically adapting to
environmental fluctuations such as atmospheric turbulence and rain
attenuation. The joint optimization of the quantum bit error rate
(QBER) and the classical signal-to-noise ratio (SNR) is formulated
as a quadratic unconstrained binary optimization (QUBO) problem, enabling
efficient adaptive phase control utilizing both quantum and classical
computational methods. Comprehensive theoretical modeling and simulations,
benchmarked against experimental data from the Micius satellite, demonstrate
substantial performance gains. Notably, the RIS-assisted system reduces
QBER from approximately 2.5\% to 0.7\%, increases the secure key rate
(SKR) to over 30,000 bits per second, and enhances classical RF SNR
by about 3 dB at high elevation angles. These results illustrate the
practical potential of   hybrid RIS-assisted satellite links
to deliver robust, efficient, and  secure global communications.
\end{abstract}

\selectlanguage{british}%
\begin{IEEEkeywords}
QKD\foreignlanguage{english}{, RIS, Satellite Communications, Quantum
Security, Dual-band Optimization. }

\selectlanguage{english}%
\end{IEEEkeywords}

\selectlanguage{english}%

\section{Introduction }

Modern wireless networks are evolving toward providing secure global
connectivity, a vision crucial for beyond-5G and 6G systems \cite{Akyildiz2020}.
Quantum Key Distribution (QKD) represents a robust security solution
resistant to quantum computing threats, relying on quantum mechanical
principles rather than classical computational complexity \cite{Hosseinidehaj2019}.
To extend QKD coverage globally, satellite-based implementations have become a promising approach due to the favorable loss characteristics of free-space optical channels over large distances \cite{Conti2024}.
An example is the Quantum Experiments at Space Scale (QUESS) mission, which successfully deployed the Micius satellite in low Earth orbit (500 km altitude) in August 2016. Using this satellite as a trusted relay, QKD keys were securely distributed between ground stations approximately 7,600 km apart \cite{Cao2022}. This achievement demonstrates the significant potential of satellites as secure quantum relays on a global scale. Importantly, these quantum links must operate simultaneously with classical communication channels. For example, traditional RF bands can be utilized for control and data transfer while quantum optical links carry encryption keys \cite{Ruiz2025}.

The main motivation for integrating classical and quantum communications into satellite systems lies in combining their distinct advantages: while optical links can offer higher theoretical capacity than RF, the key advantage of using classical RF links in satellite systems is their robustness and reliability, particularly under adverse weather conditions. On the other hand, optical free-space links are highly susceptible to atmospheric turbulence, clouds, and rain, which can severely degrade link quality or interrupt service  \cite{Kaushal2017}.  By embedding both channels into a single integrated platform, the system can simultaneously distribute secure quantum keys and efficiently transmit classical data. While the integration of quantum and classical links introduces additional hardware and control complexity at the subsystem level, it significantly reduces overall satellite payload redundancy, size, and launch costs compared to deploying separate platforms for each function. As a result, the integrated approach streamlines the system-level architecture and enhances operational efficiency, paving the way for robust, secure, and versatile global communication infrastructures. This approach represents a significant step toward the realization of globally connected, quantum-secure satellite networks \cite{Kietzmann2021,Vu2022}. By seamlessly integrating satellite-based QKD links with existing terrestrial fiber-optic infrastructures, the proposed system enhances overall communication capabilities \cite{Kietzmann2021} and enables end-to-end secure key distribution across both satellite and ground segments \cite{Vu2022}.

Despite these promising advantages, significant challenges persist
in implementing high-performance satellite-based QKD systems. Free-space quantum optical links, particularly through Earth's atmosphere, remain vulnerable to alignment issues and atmospheric disturbances such as clouds and turbulence. These factors lead to substantial signal attenuation and quantum state decoherence \cite{VazquezCastro2023}.
Traditional solutions rely on trusted relay nodes at ground stations, which act as secure intermediaries that receive, decrypt, and re-encrypt quantum keys alongside classical data. While this approach
mitigates free-space link losses and blockages, it also introduces critical security vulnerabilities, as these ground-based relays become high-value targets for adversaries \cite{Lemons2023}. These challenges emphasize the necessity for innovative techniques to enhance the reliability, robustness, and security of satellite QKD without solely depending on trusted relays.

In parallel, advancements in wireless communications have sparked significant interest in Reconfigurable Intelligent Surfaces (RIS). An RIS comprises a large number of passive, low-cost elements whose reflective properties (phase and amplitude) can be dynamically adjusted \cite{Khalil2022}. This adaptability enables RIS to effectively control wireless propagation environments, significantly enhancing coverage and interference management while avoiding the high energy consumption typical of conventional active relays or antenna arrays \cite{Ahmed2024}. Recently, RIS applications have expanded into non-terrestrial environments, including satellites and aerial platforms
\cite{Khalil2025}. For instance, RIS technology has been successfully used to improve inter-satellite Terahertz communications by mitigating
path loss and correcting beam misalignment \cite{Tekbiyik2022a}. Additionally, RIS technology has facilitated the development of hybrid free-space optical/radio-frequency (FSO/RF) downlinks by dynamically redirecting optical signals around atmospheric obstructions, while RF channels provide reliable connectivity during periods of adverse weather or optical link outages \cite{Nguyen2023}. This hybrid strategy leverages the complementary strengths of both domains, combining the high security and bandwidth of  optical channels with the robustness and continuous availability of RF links. However, existing RIS-assisted hybrid systems typically address optical and RF links independently or focus on basic switching between bands. A unified dual-band RIS architecture that enables simultaneous, real-time optimization of both quantum and classical channels is required to maximize  security and reliability in dynamically changing satellite environments.

Integrating RIS technology into quantum communication systems represents an emerging research direction. Preliminary studies have demonstrated that RIS can substantially enhance quantum network performance by creating virtual line-of-sight paths for distributing quantum entanglement, thereby significantly improving quantum state fidelity \cite{Chehimi2025}. Another innovative approach has utilized RIS as a trusted intermediary in multi-user QKD networks, securely managing classical  and quantum signals separately to reduce potential security vulnerabilities \cite{Kisseleff2023}. Despite these initial advancements, existing research has yet to comprehensively explore the integrated optimization of quantum and classical channels using RIS in satellite communication scenarios. Integrating both quantum and classical links within a unified RIS framework offers critical advantages: it reduces hardware redundancy and overall payload weight, enables synchronized adaptation to dynamic channel conditions, and improves resource utilization and security through joint channel management.  

This paper addresses a critical research gap by proposing a dual-band RIS-assisted hybrid scheme for simultaneous satellite-based quantum (850 nm optical) and classical (S-band RF) communications. In the proposed system, the quantum optical channel delivers secure key distribution via  QKD, leveraging the fundamental security guarantees of quantum mechanics. In parallel, the classical S-band RF channel ensures reliable and weather-resilient transmission of data and control signals. Central to this approach is a real-time, feedback-driven RIS phase optimization framework formulated as a Quadratic Unconstrained Binary Optimization (QUBO) problem, which dynamically adjusts RIS reflection profiles to jointly minimize the quantum bit error rate (QBER) and maximize the classical signal-to-noise ratio (SNR). This dual-band optimization strategy enables robust performance under atmospheric fading and alignment impairments, which are key challenges in free-space satellite links, while supporting the efficient and concurrent management of quantum and classical resources. The system's effectiveness is validated through benchmarking against experimental Micius satellite data, demonstrating substantial improvements in the secure key rate, SNR, and overall link robustness.

In our proposed system, we select an 850 nm optical wavelength for the quantum channel, as it aligns well with the peak sensitivity of silicon-based single-photon detectors, enabling efficient photon detection. For classical communication, we employ the S-band (2--4 GHz), which is widely used in satellite telemetry and offers reliable performance for data transmission. By integrating these two frequency bands, the system supports the independent and adaptive optimization of both quantum and classical links, thereby enhancing overall security, flexibility, and performance in hybrid satellite communication networks.

The primary contributions of this work are as follows:
\begin{itemize}
\item {We introduce a novel frequency-selective dual-band RIS-assisted satellite communication framework. This framework enables the simultaneous and independent optimization of quantum and classical communication channels, substantially extending the capabilities demonstrated in previous satellite QKD missions}.
\item {We rigorously formulate the joint quantum-classical RIS phase control problem as a QUBO, supporting efficient real-time reconfiguration via advanced classical or quantum computational methods. This approach is well-suited to dynamic satellite environments and varying atmospheric conditions}.
\item {We present comprehensive theoretical analyses and simulation results that demonstrate performance enhancements over prior benchmarks. Notable improvements include reducing the QBER to approximately 0.7\%, increasing the secure key rate (SKR) to over 30{,}000 bits/s, and enhancing the classical RF SNR by approximately 3 dB across a range of satellite elevation angles}. 
\end{itemize}

Collectively, these contributions represent a significant advancement in RIS-assisted hybrid satellite quantum-classical communications and establish a robust foundation for a secure, reliable, and efficient global quantum internet infrastructure.

The remainder of this paper is organized as follows. Section \ref{sec:2} presents the dual-channel LEO-to-ground system model, detailing the quantum (850 nm) and classical (S-band) link budgets, channel impairments, and the dual-band RIS. Section \ref{sec:3} derives a range-normalized cost function that balances QBER and SNR, formulating it as a QUBO problem solvable using various computational methods. Section \ref{sec:6} reports RIS phase assignment statistics and performance metrics benchmarked against the Micius satellite. Finally, Section \ref{sec:7} concludes  the paper and outlines directions for future research. To ensure clarity and consistency, a  summary of the principal symbols, parameters, and variable definitions used throughout the theoretical analyses is provided in Table~\ref{table-1}.

\label{table-1}\begin{table}[ht!]
\centering
\caption{Major symbols used throughout Sections 2.1--3.3}
\label{tab:major-symbols}
\begin{tabular}{ll}
\hline
\textbf{Symbol} & \textbf{Description} \\
\hline
$R_E$ & Earth's radius (approximately 6371\,km). \\
$h_{\mathrm{sat}}$ & Satellite altitude (e.g., 500--600\,km for LEO). \\
$\theta$ & Satellite elevation angle at the ground station. \\
$d$ & Slant range between satellite and ground station. \\
$d_{\mathrm{atm}}$ & Effective atmospheric path length. \\
$h_{\mathrm{atm}}$ & Effective atmospheric height (8--10\,km for optics). \\
$\lambda_Q$ & Wavelength of quantum (optical) signal (e.g., 850\,nm). \\
$\lambda_C$ & Wavelength of classical (RF) signal (e.g., 13\,cm in S-band). \\
$\kappa_Q$ & Atmospheric attenuation coeff. for optical link. \\
$\kappa_C$ & Atmospheric attenuation coeff. for RF link. \\
$H_Q$ & Baseline quantum (optical) channel gain without RIS. \\
$H_C$ & Baseline classical (RF) channel gain without RIS. \\
$H_Q^{\mathrm{RIS}},\,H_C^{\mathrm{RIS}}$ & Channel gains with RIS contributions. \\
$\Gamma_C$ & SNR (Signal-to-Noise Ratio) for the classical channel. \\
$\epsilon_Q$ & QBER (Quantum Bit Error Rate). \\
$\text{SKR}$ & Secure key rate for the quantum channel. \\
$R_{\mathrm{raw}}$ & Raw key rate before error correction. \\
$f_{\mathrm{EC}}$ & Error-correction efficiency factor. \\
$h_2(\cdot)$ & Binary entropy function. \\
$Q(x)$ & Q-function (tail probability of Gaussian). \\
$\alpha,\,\beta$ & Weights in multi-objective (QBER vs. SNR) cost function. \\
$\theta_{n}^{Q},\,\theta_{n}^{C}$ & Phase shifts (optical/RF) at RIS element $n$. \\
$b_Q,\,b_C$ & Number of phase bits for quantum/classical bands. \\
$x_{n,k}^{Q},\,x_{n,k}^{C}$ & Binary variables encoding RIS phase shifts. \\
$N$ & Number of RIS elements. \\
$\mathbf{x}$ & Vector of all binary variables in the QUBO. \\
$\mathbf{Q}$ & Quadratic coefficient matrix in the QUBO formulation. \\
$\mathbf{c}$ & Linear coefficient vector in the QUBO formulation. \\
\hline
\end{tabular}
\end{table}

\section{System Model\label{sec:2} }

In the proposed system, an LEO satellite simultaneously transmits two distinct signals to the ground: an optical quantum signal at 850 nm, used exclusively for QKD to enable secure communication, and a classical RF signal in the S-band (2--4 GHz), utilized for robust data transmission and network control. The 850 nm wavelength is selected for the quantum channel due to its compatibility with silicon-based Single-Photon Avalanche Diodes (SPADs), which exhibit peak photon-detection efficiencies between 40\% and 60\% in the near-infrared spectrum, along with low dark-count rates, facilitating practical, room-temperature quantum communication links \cite{Hosseinidehaj2019}. Additionally, this wavelength resides within an atmospheric transmission window characterized by low scattering and moderate absorption, making it suitable for satellite-based optical communication \cite{Abasifard2023}.
 The S-band is internationally allocated for space-to-ground RF communications and offers a favorable balance between rain-fade tolerance, antenna compactness, and moderate bandwidth availability, making it ideal for reliable satellite communication links \cite{Yeo2012}.  Both optical and RF signals, generated separately by onboard optical and RF transmitters, propagate along dedicated channels toward two terrestrial endpoints: a RIS and a user ground station. This dual-channel architecture supports simultaneous secure quantum communication and classical data transmission, providing essential system control, synchronization, and auxiliary data transfer.

\begin{figure}
\begin{centering}
\includegraphics[width=3.5in,viewport=2bp 0bp 550bp 350bp]{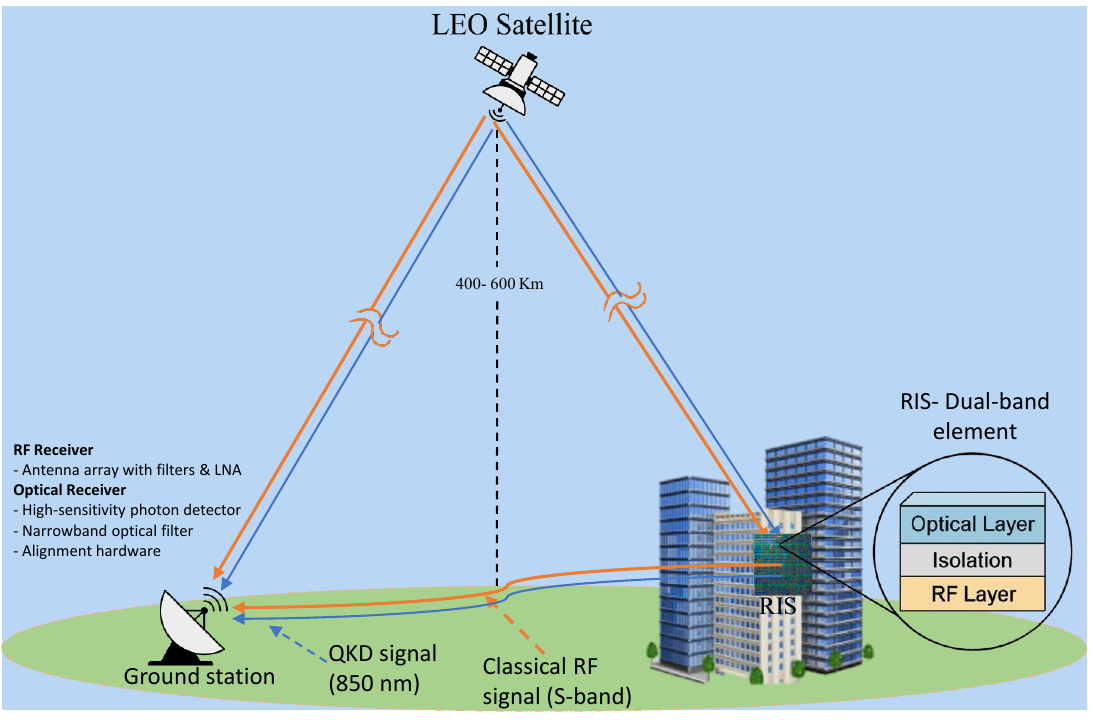} 
\par\end{centering}
\caption{System model: LEO satellite\protect\nobreakdash-to\protect\nobreakdash-ground link with dual\protect\nobreakdash-band RIS enhancement: schematic showing a 500 km\protect\nobreakdash-altitude LEO satellite transmitting simultaneous 850 nm quantum and S\protect\nobreakdash-band RF signals to a ground station, aided by a nearby dual\protect\nobreakdash-band RIS for joint QKD and classical link improvement. \label{fig:1}}
\end{figure}

The geometric configuration is depicted in Fig. \ref{fig:1} where the satellite orbits at an altitude of $h_{sat}$ $\approx400\text{\textendash}600$ km and communicates with the ground station at an elevation angle of $\theta$. The slant range between the satellite and the ground receiver is expressed as:
\begin{equation}
d=\sqrt{(R_{E}+h_{sat})^{2}-(R_{E}\cos\theta)^{2}}-R_{E}\sin\theta,
\end{equation}
with Earth's radius $\mathit{R_{E}}=6371$km. The atmospheric refraction is accounted for by the effective atmospheric path $d_{atm}:$
\begin{equation}
d_{atm}=\frac{h_{atm}}{\sin\theta+\sqrt{\sin^{2}\theta+2h_{atm}/R_{E}}},
\end{equation}
where the effective atmospheric height $h_{atm}\approx8-10$ km.

\subsection{Channel Models and Free-Space Path Loss}
Due to the significant disparity in operating wavelengths, the free-space path loss (FSPL) differs  between the optical and RF channels. We model the baseline  LoS  channel gain (i.e., without RIS assistance) for each link using the Friis transmission formula with the appropriate antenna gains \cite{Cho2019}. Specifically, for the optical channel (quantum link) we include the directivity gain of the transmitter's laser optics ($G_{t,Q}$) and the receiver telescope's aperture gain ($G_{r,Q}$), leading to an optical FSPL term $G_{t,Q}G_{r,Q}\left(\frac{\lambda_{Q}}{4\pi d}\right)^{2}$. Likewise, for the classical S-band link, we include the satellite $G_{t,C}$ and the ground antenna $G_{r,C}$ gains , yielding an RF FSPL term $G_{t,C}G_{r,C}\left(\frac{\lambda_{C}}{4\pi d}\right)^{2}$. 

For the quantum optical signal, the standard Friis transmission equation, incorporating transmitter gain $G_{t,Q}\approx4\pi/\theta_{div}^{2}$ (where $\theta_{div}$ is the beam divergence angle) and receiver telescope gain $G_{r,Q}=\pi D_{r}^{2}/\lambda_{Q}^{2}$(with telescope aperture diameter $D_{r}$), needs to be employed. The amplitude gain of the direct quantum channel $H_{Q}$, including atmospheric attenuation, turbulence-induced fading $\chi$, and pointing-induced fading $h_{pe}(t)$, can be modeled as
\begin{equation}
H_{Q}=\frac{\lambda_{Q}}{4\pi d}\sqrt{G_{t,Q}G_{r,Q}}\:e^{-\frac{\kappa_{Q}\cdot d_{atm}}{2}}\sqrt{\chi\:h_{pe}(t)},
\end{equation}
where $\lambda_{Q}=850\textrm{nm }$(optical wavelength) and $\kappa_{Q}$ ($\approx0.4-0.7$ dB/km) characterize atmospheric optical losses \cite{Miao2021}. Turbulence fading $\chi$ follows Gamma-Gamma statistics, while pointing fading $h_{pe}(t)$ accounts for residual misalignments and jitter \cite{Yang2014}.

In parallel, the RF S-band channel between the satellite and the ground station is influenced by several propagation impairments, including atmospheric attenuation, ionospheric losses, and rain fading, as discussed below.  
\begin{itemize}
\item \textbf{Atmospheric attenuation} ($L_{atm,C}$): Predominantly caused by oxygen and water vapor absorption, the attenuation is modeled as $L_{atm,C} = \exp(-\kappa_C \cdot d_{atm})$, where $\kappa_C \approx 0.01\text{--}0.05$ dB/km is the atmospheric attenuation coefficient and $d_{atm}$ is the effective atmospheric path length~\cite{Pratt2019}.
\item \textbf{Ionospheric losses} ($L_{ion}$): These losses depend on the total electron content (TEC) and scintillation index ($S_4$), and are given by
\begin{equation}
L_{ion}=10^{-I_{ion}/10},I_{ion}=0.0265\cdot\frac{\text{TEC}}{f_{C}^{2}}+0.018\cdot S_{4}\cdot\frac{f_{ref}^{1.5}}{f_{C}^{1.5}},\label{eq: LL}
\end{equation}
where $f_{C}$ is the carrier frequency (GHz), $\text{TEC} \approx 5\text{--}50$ TECU, $S_{4} \approx 0.1\text{--}0.5$, and $f_{ref} = 1$ GHz~\cite{Ippolito1999}.
\item \textbf{Rain fading} ($L_{rain}$): At S-band, rain-induced attenuation is generally modest, modeled as $L_{rain} = \exp(-\gamma_{R}$\,$ d_{rain})$, with $\gamma_{R} = k R^{\alpha}$, where $R$ is the rain rate (mm/h), and  $10^{-4} \leq k \leq 5 \times 10^{-4}$, $\alpha \approx 1.0\text{--}1.2$ are frequency-dependent empirical parameters~\cite{Jr.2017}.
\end{itemize}

\noindent
Taking these effects into account, the overall amplitude gain of the S-band RF channel is expressed as~\cite{Seybold2005}:
\begin{equation}
H_{C} = \frac{\lambda_{C}}{4\pi d} \sqrt{G_{t,C} G_{r,C}} \sqrt{L_{atm,C} \cdot L_{ion} \cdot L_{rain}},
\end{equation}
where $\lambda_{C} \approx 10\text{--}15$ cm is the RF wavelength, and $d$ is the slant range.

\subsection{Dual-Band RIS Phase Shift Model \label{sec:4}}

The considered RIS architecture is frequency-selective and operates simultaneously in two  bands: an optical quantum channel around 850\,nm, or 353\,THz, and a classical RF channel in the S-band (2.3\,GHz). In contrast to traditional satellite QKD systems, which typically employ separate physical channels for quantum and classical links \cite{Liao2017}, the dual-band RIS (shown in the inset of Fig. \ref{fig:1}) provides a unified, shared-aperture platform for the concurrent and independent optimization of both channels. This design allows  the simultaneous enhancement of QBER and SNR in the quantum and classical links, respectively, through programmable phase control at each band \cite{Tsafaras2025}. 

In the proposed system model, the dual-band RIS is represented as an ideal metasurface that enables independent phase manipulation for each band within the same physical aperture. Specifically, for a dual-band RIS consisting of $N$ unit cells, each element is characterized by two independent phase shift parameters:
\begin{equation}
\theta_{n}^{Q}\in[0,2\pi),\label{eq: f_Q}
\end{equation}
for the quantum optical band, and
\begin{equation}
\theta_{n}^{C}\in[0,2\pi),\label{eq:f_C}
\end{equation}
for the classical (RF) band, where $n=1,\ldots,N$.

The overall RIS-assisted channel model is described as follows: 
\begin{equation}
\Gamma(f)=A(f)e^{j\phi(f)},
\end{equation}
where $A(f)$ and $\phi(f)$ denote the amplitude and phase functions corresponding to the operating frequency $f$. Independent phase shift variables are defined at the quantum frequency $f_{Q}=c/\lambda_{Q}\approx353\mathrm{THz}$ and the classical frequency $f_{C}=c/\lambda_{C}\approx2.3\mathrm{\:GHz}$. The amplitude response $A(f)$ is assumed to be unity (lossless RIS assumption) in each band for theoretical tractability.

 The central modeling assumption in this work is that the phase shifts at each band can be controlled independently. Specifically, the adjustment of the quantum channel phase, $\theta_{n}^{Q}$, does not influence the classical channel phase, $\theta_{n}^{C}$, and vice versa. As a result, any electromagnetic cross-coupling between the quantum and classical phase control channels is considered negligible and is not incorporated into the model. This assumption allows for the independent optimization of the QBER, through ${\theta_{n}^{Q}}$, and the SNR, through ${\theta_{n}^{C}}$, over the same RIS aperture.
 
It is important to highlight that recent reviews and studies on dual-band and frequency-selective metasurfaces \cite{Chen2024,Fontoura2021} have explored multiband  Frequency Selective Surface (FSS) and RIS architectures for wireless and satellite applications. However, these works are generally limited to either band-switching or sequential (non-concurrent) optimization. In other words, they do not provide a framework for the truly simultaneous and independent optimization of two distinct communication channels, such as quantum and classical links, using a single, unified RIS platform.

By contrast, the dual-band RIS model proposed in this work enables, for the first time, fully concurrent and independently programmable phase control for both quantum (optical) and classical (RF) channels within the same metasurface. This unique capability allows for the simultaneous, real-time optimization of both the  QBER  and the SNR  over a shared RIS aperture. Such an approach directly addresses key open challenges identified in contemporary literature and represents a significant advancement in the design and application of frequency-selective metasurface technologies.

\subsection{RIS-Assisted Channel Model and Phase Quantization}
In the system, a dual-band RIS consisting of $N$ reflective elements is installed at a suitable ground location, such as the rooftop of a tall building, to enhance both quantum and RF channels through coherent reflection. Each RIS element independently imparts discrete and quantized phase shifts ($\theta_{n}^{Q}$, $\theta_{n}^{C}$) onto the incident quantum and classical signals, respectively, defined as
\begin{equation}
\theta_{n}^{Q}=\frac{2\pi}{2^{b_{Q}}}\sum_{k=0}^{b_{Q}-1}2^{k}x_{n,k}^{Q},\:\quad\theta_{n}^{C}=\frac{2\pi}{2^{b_{C}}}\sum_{k=0}^{b_{C}-1}2^{k}x_{n,k}^{C},\label{eq: ThQ_ThC}
\end{equation}
where $b_{Q}$ and $b_{C}$ are the phase quantization resolutions (in bits) for the quantum and classical domains, respectively, and $x_{n,k}^{Q},\:x_{n,k}^{C}\in\left\{ 0,1\right\} $ represents the bit states for the phase settings of each RIS element.

The satellite-to-ground communication occurs over two primary propagation paths: the direct LoS path and the RIS-assisted indirect path. The combined amplitude of the RIS-assisted quantum channel at the receiver is:
\begin{equation}
H_{Q}^{tot}=H_{Q}+\sum_{n=1}^{N}H_{Q,S\rightarrow RIS,n}\cdot e^{j\theta_{n}^{Q}}\cdot H_{Q,RIS\rightarrow G,n},\label{eq: Hq}
\end{equation}
Similarly, the combined classical RF channel is expressed as follows:
\begin{equation}
H_{C}^{tot}=H_{C}+\sum_{n=1}^{N}H_{C,S\rightarrow RIS,n}\cdot e^{j\theta_{n}^{C}}\cdot H_{C,RIS\rightarrow G,n},\label{eq:Hc}
\end{equation}
where $H_{\mathtt{X}-RIS,n}$ and $H_{\mathtt{X},RIS-G,n},$ denote the sub-path amplitude gains from the satellite to the RIS and from the RIS to the ground station, respectively.

\subsection{Receiver Performance Metrics}
The ground station is equipped with dual-band receivers designed for the simultaneous and independent reception of optical and classical RF signals. For the quantum optical channel, a high-sensitivity photon detector featuring narrowband optical filtering and precise beam alignment mechanisms ensures effective quantum signal detection with minimal noise and background interference. Conversely, the classical RF signals are received by dedicated antenna arrays that incorporate precise RF filtering and amplification stages, thereby maintaining high signal integrity.

\subsubsection{RF Channel Performance Metrics}
For the classical RF link, the critical performance metrics include the received signal-to-noise ratio (SNR, denoted as ($\Gamma_{C}$) and the corresponding bit error rate (BER). The SNR for the RF link is defined as:
\begin{equation}
\Gamma_{C}=\frac{P_{t,C}G_{t,C}G_{r,C}|H_{C}^{tot}|^{2}}{k_{B}T_{sys}B}
\end{equation}
where $P_{t,C}$ is the RF transmit power, $k_{B}$ is Boltzmann's constant, $T_{sys}$ is the system noise temperature, and $B$ is the bandwidth of the RF signal. 

Due to its standardization in DVB-S2 systems~\cite{Morello2006} and widespread adoption in recent LEO broadband architectures~\cite{Su2019}, we consider QPSK modulation, whose constant-envelope waveform enables satellite power amplifiers to operate efficiently near saturation. Its two-bit-per-symbol mapping also doubles the spectral efficiency of BPSK without requiring additional bandwidth. Under additive white Gaussian noise, the QPSK bit-error rate (BER) is given by
\begin{equation}
\text{BER} \simeq Q\left(\sqrt{2\Gamma_{C}}\right).
\end{equation}
In the proposed scenario, multipath propagation is negligible due to the highly directive RF antennas employed, which possess narrow beamwidths that suppress ground-level scatterers and multipath components in satellite-to-ground links~\cite{You2000}.

\subsubsection{Quantum Optical Channel Performance Metrics}
The performance of the quantum optical link is quantified by two critical metrics:QBER, denoted by $\epsilon_{Q}$) and Secure Key Rate (SKR). The QBER directly impacts the fidelity of quantum key distribution and accounts for factors such as turbulence-induced fading, atmospheric loss, pointing errors, and detector imperfections. Mathematically, the QBER is formulated as \cite{Andrews2005}:
\begin{equation}
\epsilon_{Q}=\frac{1}{2}\left(1-V(\chi)\;|H_{Q}^{tot}|^{2}\right)+p_{dark},
\end{equation}
where the visibility function $V(\chi),$ impacted by turbulence is expressed as follows:
\begin{equation}
V(\chi)=V_{0}\cdot\exp^{-\frac{\sigma_{\phi}^{2}}{2}}.
\end{equation}

Here, $V_{0}\in\left[0.9,0.98\right]$ is the baseline interferometric
visibility, and $\sigma_{\phi}^{2}\thickapprox1.03$ is the phase variance induced by atmospheric turbulence (related to the Rytov variance). Additionally, $p_{dark}$ accounts for dark count probabilities in photon detectors, typically ranging between $10^{-6}$ and $10^{-5}$ \cite{Tatarski1961}.

The SKR, on the other hand, quantifies the achievable
secure information rate while accounting for errors and practical coding inefficiencies. The SKR expression is given by \cite{Hosseinidehaj2019}:
\begin{equation}
\text{SKR}=R_{raw}\,[1-2h_{2}(\epsilon_{Q})]-f_{EC}\,R_{raw}\,h_{2}(\epsilon_{Q}),
\end{equation}
where $R_{raw}\propto\mid H_{Q}\mid^{2}$ is the raw key generation rate, $h_{2}(x)$ is the binary entropy function that quantifies information loss due to errors, and $f_{EC}\in\left[1.1,1.2\right]$ accounts for error correction inefficiency. The SKR typically decreases sharply at lower elevation angles due to increased atmospheric path loss and turbulence-induced errors, as experimentally demonstrated during the Micius mission \cite{Liao2017}.

\section{Joint Optimization and QUBO Formulation\label{sec:3}}

In the preceding discussion, a dual-channel satellite-to-ground communication system was introduced, wherein the quantum optical link, operating around 850\,nm for QKD, and the classical RF link, in the S-band, coexist to provide secure key distribution and robust classical connectivity. The system model accounted for channel gains in both optical and RF domains, incorporating free-space path loss, atmospheric attenuation, turbulence effects, and RIS-assisted propagation. The quantum bit error rate $\epsilon_{Q}$ and the classical SNR $\Gamma_{C}$ were identified as key performance measures for these two links. Minimizing $\epsilon_{Q}$ is crucial for sustaining high-fidelity QKD, while maximizing $\Gamma_{C}$ ensures reliable data rates and link robustness. This section formally establishes a joint optimization framework that merges these objectives and subsequently formulates the resulting problem as a  QUBO model.
\subsection{Joint Optimization Problem}

The dual-band RIS must ensure that the QBER remains below the 11\% security threshold established by the BB84 protocol \cite{Shu2023} for the quantum (optical) channel while also guaranteeing that the classical SNR is sufficiently high for reliable data decoding in the RF band. A value of 20 dB (i.e., $\Gamma_{C} \approx 100$ in linear units) is widely regarded as a practical SNR margin for reliable data transmission in classical satellite communication systems \cite{Tomaello2011}.
Let the discrete phase shifts applied by the $n^{th}$ RIS element be $\theta_{n}^{Q}$ for the quantum channel and $\theta_{n}^{C}$ for the classical channel. The resulting multi-objective problem can be stated as:
\begin{gather}
\min_{\{\theta_{n}^{Q},\theta_{n}^{C}\}}\quad\epsilon_{Q},\max_{\{\theta_{n}^{Q},\theta_{n}^{C}\}}\quad\Gamma_{C} \nonumber \\
\text{s.t.}\quad\theta_{n}^{Q},\theta_{n}^{C}\in[0,2\pi),\quad\forall n\in\{1,\ldots,N\}\label{eq: min_Max}
\end{gather}

To obtain a tractable optimization problem, these objectives are merged into a single scalar cost function using the weighted-sum method, which is a standard approach in multi-objective optimization:
\begin{equation}
F(\{\theta_{n}^{Q},\theta_{n}^{C}\})=\alpha\epsilon_{Q}-\beta\log_{2}\left (1+\Gamma_{C}\right).\label{eq: F(f)}
\end{equation}
where $\alpha$ and $\beta$ are non-negative weighting coefficients. 

To ensure that the cost function in \eqref{eq: F(f)} provides balanced sensitivity to both the QBER and classical channel spectral efficiency, the weights $\alpha$ and $\beta$ must be  normalized to account for their differing scales. Typically, $\epsilon_{Q}$ for satellite quantum links lies in the range $10^{-3}$ to $10^{-2}$, while the classical spectral efficiency $\log_{2}\left(1+\Gamma_{C}\right)$ ranges from approximately 3 to 8 for SNRs between 20 and 30 dB \cite{Sidhu2022,Li2019a}. Therefore, directly adding these terms results in the SNR contribution dominating the cost function. 

To correct this imbalance, a range normalization procedure is employed \cite{Kaddani2017}. Specifically, the weights are chosen such that each term contributes equally to the cost near their operational thresholds. The weights are defined as: $\alpha=1$ and $\beta=\epsilon^{\star}/\log_{2}\left(1+\Gamma_{C}^{\star}\right)$,
where $\epsilon^{\star}=0.011$ is the BB84 security threshold (adjusted for practical implementation losses) \cite{Shor2000}, and  $\Gamma_{C}^{\star}=100$ (i.e.,$\,20$\,dB) is a representative SNR target for secure satellite downlinks. Substituting these values gives $\beta\approx1.65\times10^{-3}$. The resulting cost function is
\begin{equation}
F(\theta)=\epsilon_{Q}(\theta)-1.65\times10^{-3}\log_{2}\left(1+\Gamma_{C}(\theta)\right),\label{eq: F(0)}
\end{equation}
where $\theta=\left\{ \theta_{n}^{Q},\theta_{n}^{C}\right\} _{n=1}^{N}$ denotes the RIS phase vector for all elements in both frequency bands.

In this formulation, minimizing $F$ penalizes high QBER and rewards high SNR in a balanced manner. Reducing $\epsilon_{Q}$ or increasing $\Gamma_{C}$ makes $F$ more negative, which aligns with the Pareto knee-point criterion for balanced performance \cite{Zheng2023}.

For rapidly varying channel conditions, a dynamic normalization (\textquotedblleft swing-weight \textquotedblright) scheme \cite{Parnell2009} can be adopted:
\begin{equation}
\alpha=\frac{1}{\epsilon^{\star}},\quad\beta=\beta_{o}\frac{\log_{2}\left(1+\Gamma_{C}^{\star}\right)}{\log_{2}\left(1+\Gamma_{C}\right)},\qquad\beta_{o}=0.01,
\end{equation}
so that the cost function remains balanced as channel conditions fluctuate.
\subsection{QUBO Model Formulation}

Building on the weighted cost function $F$ defined in \eqref{eq: F(0)}, we now formulate the finite-resolution phase selection problem as a QUBO model. This formulation enables the application of both emerging quantum annealing hardware and advanced classical heuristics to determine the optimal RIS phase configuration for dual-band operation.

Each RIS element, indexed by $n\in\{1,2,...,N\}$,  supports a discrete set of phase shifts: $2^{b_{Q}}$ levels for the quantum optical band and $2^{b_{C}}$ levels for the classical RF band. To represent these discrete phase levels, we define the following binary variables:
\begin{gather}
x_{n,k}^{Q},\:x_{n,k}^{C}\in\{0,1\},\qquad\quad k=0,...\dots,b_{Q}-1\:\textrm{or}\:b_{C}-1\nonumber \\
\textrm{s.t}.\quad\theta_{n}^{Q}=\frac{2\pi}{2^{b_{Q}}}\sum_{k=0}^{b_{Q}-1}2^{k}x_{n,k}^{Q},\quad\theta_{n}^{C}=\frac{2\pi}{2^{b_{C}}}\sum_{k=0}^{b_{C}-1}2^{k}x_{n,k}^{C},\label{eq:bb}
\end{gather}

\begin{equation}
\mathbf{x}=\left[x_{1,0}^{Q},\ldots,x_{N,b_{Q}-1}^{Q},\,x_{1,0}^{C},\ldots,x_{N,b_{C}-1}^{C}\right]^{\mathrm{T}}\in\{0,1\}^{N_{b}},
\end{equation}
where $N_{b}=N(b_{Q}+b_{C})$ denotes the total number of binary decision variables, i.e., the number of bits required to specify all discrete phase assignments for all RIS elements in both the quantum (optical) and classical (RF) domains. Each block of $\ensuremath{b_{Q}}($or $\ensuremath{b_{C}}$) variables, $\{x_{n,k}^{Q}\}$ (or $\{x_{n,k}^{C}\}$), uniquely encodes the quantized phase shift for the $n$$^{th}$ element in the corresponding band. Thus, the vector $\mathbf{x}$ defines a specific system-wide phase configuration, and the optimizer seeks the binary vector that minimizes the joint performance cost function under hardware constraints. 

By substituting the quantized phase vectors $\theta_{n}^{Q}$ and $\theta_{n}^{C}$ into the composite quantum and classical channel gain models $H_{Q}^{tot}$ and $H_{C}^{tot}$, respectively, the entire RIS configuration is determined by the binary vector \textbf{x}. Thus, the system performance metrics, such as QBER and SNR, become explicit functions of \textbf{x}, and the optimization problem can be formulated directly in terms of this binary vector.

Upon expanding the squared magnitudes of the composite channel gains using Euler's identity, cross terms of the form $\cos\left(\theta_{m}^{Q}-\theta_{n}^{Q}\right)$ and $\cos\left(\theta_{m}^{C}-\theta_{n}^{C}\right)$ naturally arise. These cosine terms are inherently nonlinear with respect to the underlying
binary variables in \textbf{x},\textbf{ }since each phase $\theta_{n}^{Q}$ and $\theta_{n}^{C}$  is represented as a weighted sum of binary decision variables. To ensure that the optimization problem remains in the standard QUBO form, i.e., a quadratic objective function in the binary variables, we approximate each cosine term using a second-order Taylor expansion: $cos(\Delta\theta)\approx1-\frac{1}{2}(\Delta\theta)^{2}$,
where $\Delta\theta$ is the phase difference between any two elements in the respective band. This transformation enables each cross term to be represented as a quadratic polynomial in the binary vector \textbf{x}. The approximation error, denoted $\epsilon_{apx}$, remains well-controlled and is typically limited to a few percent for practical quantization resolutions, such as 2-bit phase control (corresponding to 11.25° steps)\cite{Glover2018}. 

By collecting all quadratic and linear terms arising from both the quantum and classical domains, as well as the constant offsets inherited from the weighted cost function, the resulting QUBO objective function can be expressed succinctly as:
\begin{equation}
\mathrm{\underset{\mathit{x}\in\{0,1\}^{\mathit{N}_{b}}}{min}\quad}F(x)=x^{T}Qx+c^{T}x, \label{eq: F(x)}
\end{equation}
where $Q\in\mathbb{R^{\mathrm{N_{b}\times N_{b}}}}$ encodes pairwise
interactions arising from both direct and RIS-reflected signal paths,
including inter-band coupling effects, and $c\in\mathbb{R^{\mathrm{N_{b}}}}$ aggregates all linear coefficients, encompassing contributions from self-interference, detector noise, and calibration offsets. 

For a representative satellite downlink with $N=100$ elements and $b_{Q}=b_{C}=2$ bits of phase resolution, the total number of binary variables is $N_{b}=400$. Numerical evaluations confirm that the truncated quadratic model remains well-conditioned, with the worst-case objective deviations caused by the second-order cosine approximation staying below 2\% for the typical 2-bit phase quantization.

This QUBO formulation can be addressed directly on quantum annealers (e.g., D-Wave systems) or through gate-model hybrid algorithms such as the Quantum Approximate Optimization Algorithm (QAOA), which effectively minimizes binary-quadratic objectives \cite{Lechner2020}. Alternatively, classical metaheuristics such as tabu search, simulated annealing, or block-coordinate descent can find high-quality solutions within seconds for problem sizes involving a few hundred binary variables \cite{Gokhale2020}. The resulting binary vector $x^{\star}$ directly maps to the phase settings $\{\theta_{n}^{Q\star},\theta_{n}^{C\star}\}$ that concurrently minimize QBER and maximize SNR under the hardware's discrete-phase constraints.
\section{Results and Discussions\label{sec:6} }

In this section, we present the numerical simulation results generated using the system parameters specified in Table ~\ref{tab:params}. The performance evaluation includes the QBER, SKR, SNR, and a composite cost function, each examined across a range of satellite elevation angles. Two configurations are analyzed:
\begin{enumerate}
\item the baseline unassisted case, corresponding to the conventional dual-band satellite-to-ground communication link without deployment of a  RIS; and
\item the RIS-assisted case, representing the proposed scheme in which a dual-band RIS is employed to enhance the communication link according to the developed communication-theoretic framework and optimization strategies introduced in Sections \ref{sec:2} and \ref{sec:3}.
\end{enumerate}

To benchmark the effectiveness of the proposed RIS-assisted system, the simulated performance metrics, including QBER, SKR, and SNR, are compared with publicly available experimental data from the Micius satellite mission. All simulation outcomes are obtained using an analytical model with idealized RIS parameters, representing theoretical upper-bound scenarios rather than measurements limited by practical hardware constraints~\cite{Wang2013}. This approach enables a comprehensive evaluation of the potential performance improvements achievable through RIS integration and facilitates the validation of the proposed model against established experimental benchmarks.
\begin{table}[h]
\caption{Simulation Parameters}
\label{tab:params}
\centering
\begin{tabular}{ll}
\toprule
\textbf{Parameter}               & \textbf{Value}                \\
\midrule
Satellite altitude               & 500 km                        \\
Quantum wavelength               & 850 nm                        \\
Classical wavelength             & 15 cm (S-band)                \\
Number of RIS elements           & 100--512                          \\
RIS phase resolution (optical/RF)& 2 bits each                   \\
Atmospheric attenuation ($\alpha_{Q}$) & 0.046 /km (linear)$^{\dagger}$ \\
Atmospheric attenuation ($\alpha_{C}$) & 0.0046 /km (linear)$^{\dagger}$ \\
Refractive index structure       & $5 \times 10^{-14}\ \mathrm{m}^{-2/3}$ \\
Pointing error jitter            & 2 $\mu$rad                    \\
RF transmit power                & 10 W                          \\
System noise temperature         & 290 K                         \\
Bandwidth                        & 100 MHz                       \\
Target QBER                      & 0.05                          \\
Target classical SNR             & 10 dB                         \\
\bottomrule
\end{tabular}
\vspace{1mm}
\\
\footnotesize
$^{\dagger}$For example, if originally specified as 0.2 dB/km, convert via $\alpha=0.2/(10\ln 10)\approx 0.046/\text{km}$.
\end{table}



\begin{figure}[!t]
  \centering
  \includegraphics[width=3.5in]{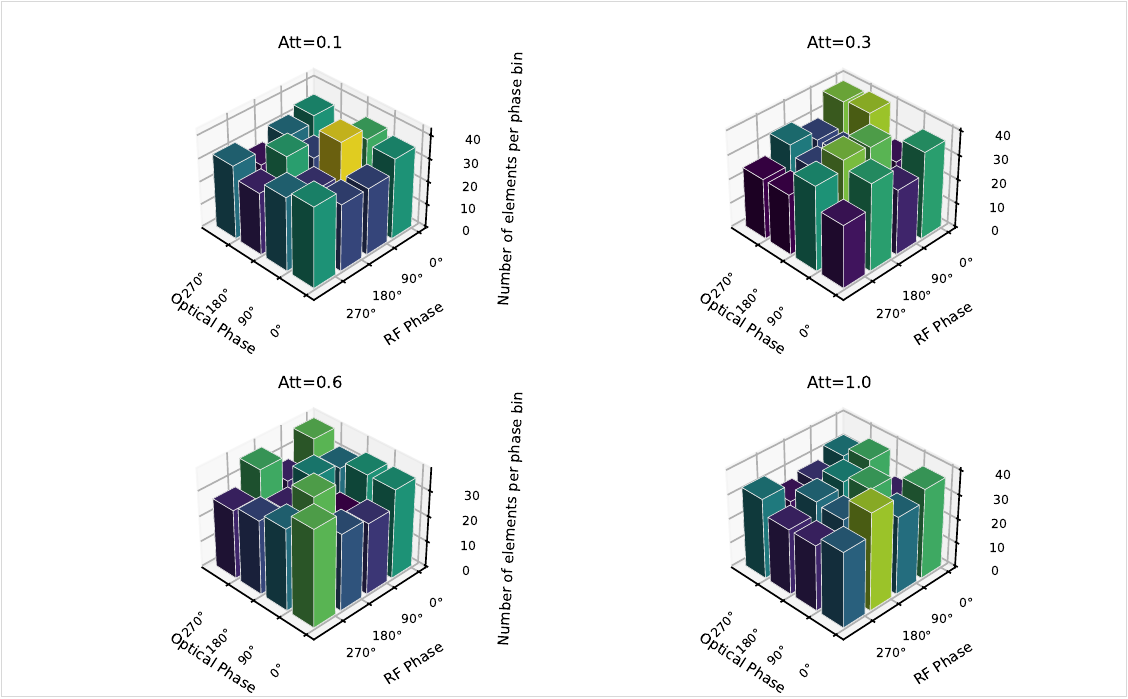}
  \caption{3D joint histograms of dual-band RIS phase assignments for varying attenuation (Att) levels. Each bar shows the number of RIS elements assigned to a particular pair of optical (quantum) and RF (classical) discrete phase bins, illustrating the system’s capacity for simultaneous and independent phase optimization under practical impairments and strict quantum security constraints.}
  \label{fig:4}
\end{figure}

To illustrate the independent phase manipulation enabled by the dual-band RIS, Fig.~\ref{fig:4} presents joint three-dimensional histograms of the optimized discrete-phase assignments for all \(N = 512\) RIS elements across both the optical (quantum) and RF (classical) bands.  Results are shown for four representative values of the unified attenuation factor $\ensuremath{\mathrm{Att}}\ensuremath{\in\{1.0,\,0.6,\,0.3,\,0.1\}}$.  Here, this single linear-scale parameter aggregates all deterministic path losses and is defined by
 $\mathrm{Att}=\left(\frac{\lambda_{X}}{4\pi d}\right)^{2}\,e^{-\kappa\,d_{\mathrm{atm}}}\,L_{\mathrm{ion}}\,L_{\mathrm{rain}},\in(0,1],$ where the free-space path loss term and the atmospheric factor \(e^{-\kappa d_{\mathrm{atm}}}\) apply in both bands, while \(L_{\mathrm{ion}}\) and \(L_{\mathrm{rain}}\) are unity for the optical link and follow Eq. \eqref{eq: LL} for the RF link.  

For each attenuation scenario, the RIS phase shifts are determined independently for every element through a joint optimization process that seeks to simultaneously minimize the QBER for the optical channel and maximize the SNR for the classical (RF) channel. During this optimization, a strict quantum security constraint is imposed: only those phase assignments that result in a QBER not exceeding the BB84 protocol threshold ($\epsilon_{Q}\leq0.11$) are considered valid solutions. Any configuration that produces a QBER above this threshold is systematically excluded from the set of feasible RIS phase combinations. 

The resulting histograms demonstrate that the RIS achieves fully independent and adaptive phase control in both the optical and RF bands. This is evidenced by the fact that, for each attenuation scenario, all possible pairings of discrete optical and RF phase assignments are utilized across the RIS elements. Since each band uses 2-bit quantization (allowing four distinct phase values: 0°, 90°, 180°, and 270°), each RIS element can be assigned any combination of these phases in the optical and RF domains. In total, this yields sixteen possible (optical, RF) phase pairs, and the presence of every combination in the histograms indicates a true independent phase assignment capability for each element. As the attenuation factor \( \text{Att} \) decreases, corresponding to a larger total path loss, the distribution of phase assignments becomes more uniform. This trend reflects that, under stronger attenuation, the ability of the RIS to align phases constructively is diminished. Importantly, all accepted phase configurations strictly satisfy the quantum security constraint, ensuring secure quantum communication at all times. These results collectively highlight the practical robustness and flexibility of the dual-band RIS architecture in enabling simultaneous quantum-secure and high-SNR classical communication, even in the presence of significant channel impairments.

\begin{figure}
\centering{}\includegraphics[width=3.5in]{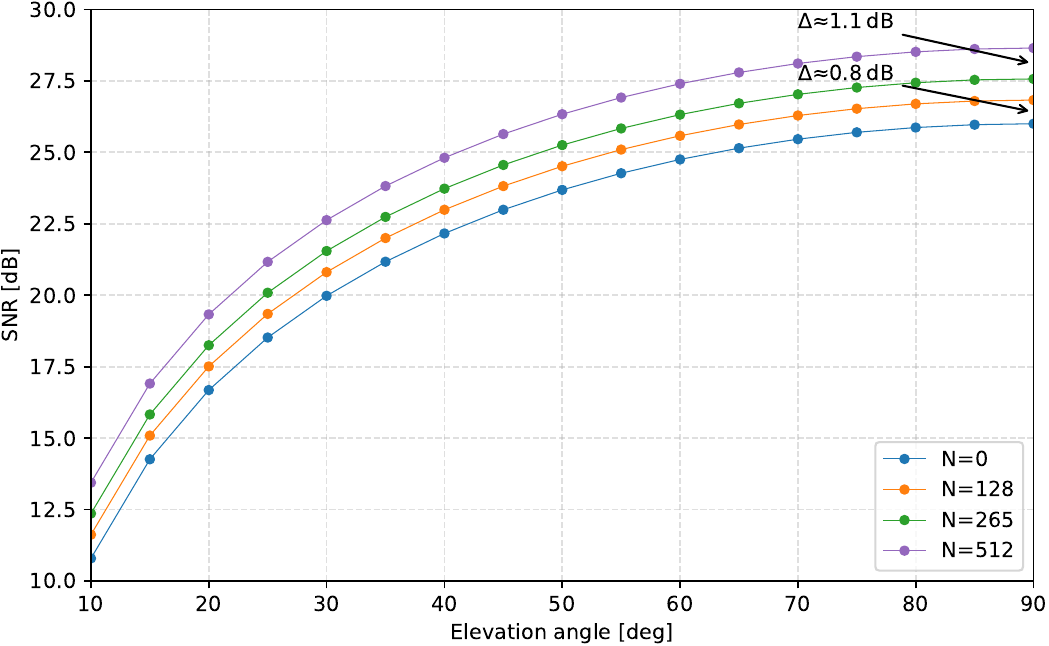}
\caption{Classical S-band SNR versus satellite elevation angle for baseline ($N\,=\,0$) and RIS-assisted links ($N\,=\,128,265,512$). SNR increases with both elevation and RIS array size; however, the incremental SNR improvement, quantified as $\Delta SNR(\mathit{N})=SNR(\mathit{N})-SNR(0)$, plateaus beyond $N\,\approx\,500$ due to practical limitations such as mutual coupling and 2-bit phase quantization, which limit further coherent combining gains.}
\label{fig:5}
\end{figure}

Fig. \ref{fig:5} shows the classical S-band SNR as a function of satellite elevation angle for several RIS configurations $N\in$$\left\{ 0,128,256,512\right\} $ elements. As expected, for the baseline case ($N=0$), the SNR increases from approximately 11 dB at a $10{^\circ}$ elevation to about 26 dB at zenith, reflecting the decrease in free-space and atmospheric path loss as elevation angles increase. Introducing a dual-band RIS significantly enhances the SNR at all elevations, with larger arrays providing greater improvements. The incremental SNR gain, $\mathrm{\Delta SNR(\mathit{N})=SNR(\mathit{N})-SNR(0)}$, quantifies the benefit of RIS deployment over the baseline. For example, at high elevations, the RIS yields SNR improvements of about 0.8\,dB for $N=128$ elements and up to 1.1\,dB for $N=512$ elements. This SNR gain directly translates to improved link robustness and throughput, especially in scenarios with lower elevation angles where the baseline SNR is weakest.

The dual-band RIS architecture enables classical channel enhancements without disturbing the quantum channel. Importantly, while increasing the number of RIS elements enhances the SNR, the marginal improvement decreases for larger arrays due to practical effects such as mutual coupling, finite phase quantization, and limited physical aperture. As a result, the observed SNR scaling saturates beyond a few hundred elements, deviating from the ideal quadratic growth expected for unconstrained array gains. These findings demonstrate that the dual-band RIS can provide meaningful and independent performance improvements for classical satellite downlinks while considering realistic hardware constraints.

\begin{figure}[!t]
  \centering
  \includegraphics[width=3.5in]{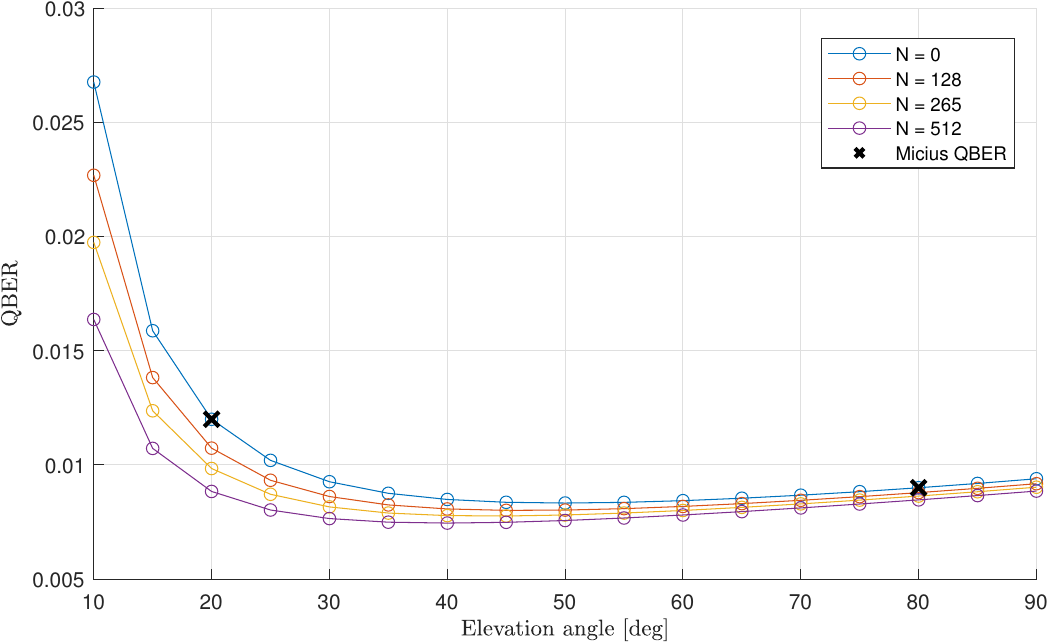}
  \caption{Impact of elevation angle on QBER in RIS-enhanced satellite QKD.}
  \label{fig:6}
\end{figure}

Turning now to quantum performance, Fig. \ref{fig:6} compares the QBER versus satellite elevation angle for a baseline link ($N\,=\,0$) and RIS-assisted configurations with $N\,=\,128,265$, and 512 elements. Crucially, our model aligns closely with the experimental QBER measurements from the Micius satellite,which is approximately 1.2\,\% at 20° elevation, as marked on the figure. This agreement validates the accuracy of our atmospheric turbulence and pointing-error modeling \cite{Liao2017,DiRenzo2020}.

It can be seen that deploying a dual-band RIS significantly improves quantum performance. With 128 elements, QBER at 20° drops to  1.02\,\%, with 265 elements further reducing it to  0.98\,\%, and with 512 elements reaching as low as  0.75\,\%, while maintaining $\leq0.72\%$\, at higher elevations. These results exceed both the baseline and the Micius benchmarks (without RIS), demonstrating the RIS's potential to enhance satellite-based QKD. The QBER improvement is quantified by $\Delta QBER(\mathit{N})=QBER(\mathit{0})-QBER(N)$, which grows with $N$ but saturates as practical system limitations, such as 2-bit phase quantization, mutual coupling, and finite aperture, become dominant. Consequently, the observed gains deviate from the ideal $\propto N^{2}$ scaling predicted by theoretical RIS models. This saturation behavior has been corroborated by recent studies on RIS-enhanced free-space quantum channels 
\cite{Chehimi2025}. Thus, the results in Fig. \ref{fig:6}, in addition to confirming the consistency of our model with real-world satellite QKD data, clearly demonstrate the advantages of integrating a dual-band RIS. The deployment of a properly engineered RIS leads to substantial reductions in QBER across all elevation angles. In particular, large-scale RIS arrays significantly enhance both the performance and reliability of satellite-based QKD.

\label{T2}\begin{table}[ht]
\centering
\footnotesize
\caption{QBER Reduction at $\theta  = 20^\circ$ and $80^\circ$ for Different RIS Sizes}
\label{tab:qber_theta}
\begin{tabular}{c c c c}
\toprule
$\boldsymbol{\theta}$ [deg]& \makecell{Baseline (N=0)\\$\equiv$ Micius 2017} & \makecell{QBER\\(N=128 / 265 / 512)} & \makecell{Reduction\\vs. Baseline} \\
\midrule
\textbf{20$^\circ$} & 1.20\,\% & 1.02\,\% / 0.98\,\% / 0.75\,\% & 15\,\% / 18\,\% / 37\,\% \\
\textbf{80$^\circ$} & 0.90\,\% & 0.84\,\% / 0.77\,\% / 0.71\,\% & 7\,\% / 14\,\% / 21\,\% \\
\bottomrule
\end{tabular}
\end{table}

\begin{figure}
\centering{}\includegraphics[width=3.5in]{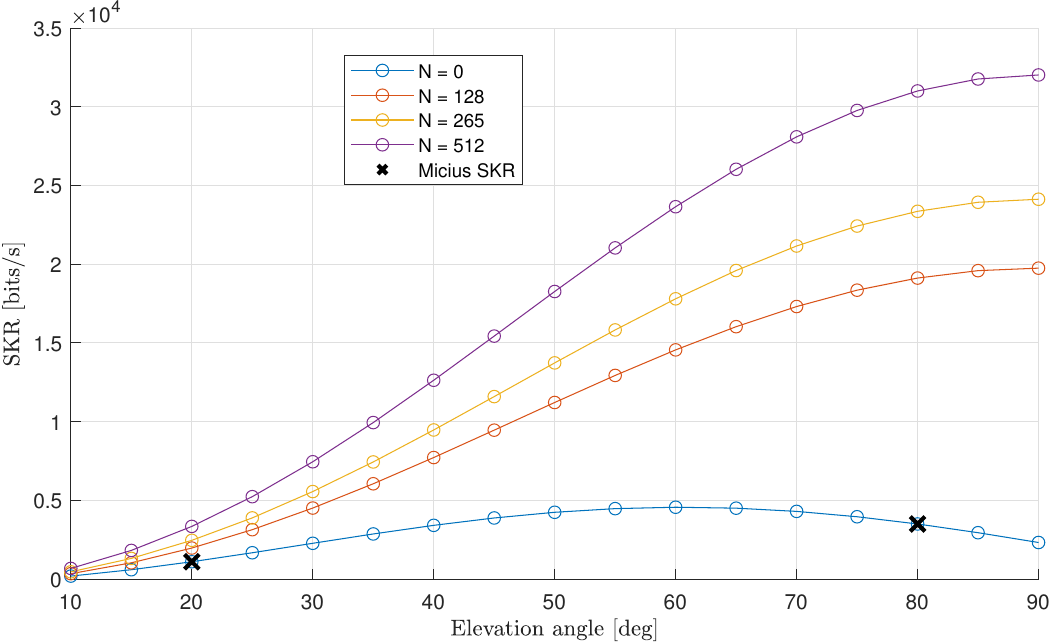} \caption{SKR performance across elevation for baseline and RIS configured links
\label{fig: 7}}
\end{figure}

\selectlanguage{english}%
Further examination of the quantum link enhancement observed in Fig. \ref{fig:6},
Fig. \ref{fig: 7} illustrates the SKR as a function of the satellite elevation angle. Without RIS (baseline), SKR peaks modestly at around 4,000 bits/s at higher elevations. The deployment of RIS significantly improves SKR performance across all elevations. With a 128-element RIS, SKR is enhanced by approximately $25\%$, further increasing with larger arrays. Specifically, the 265-element RIS delivers about a $53\%$ SKR improvement, while the largest RIS configuration (512 elements) demonstrates over $100\%$ enhancement, achieving SKRs exceeding 30,000 bits/s at zenith.  The accompanying table \ref{tab:SKR_Enhancement} summarizes these substantial performance gains compared to the baseline  Micius satellite benchmarks, highlighting the clear advantages of RIS integration.

\begin{table}[htbp]
\centering
\caption{SKR enhancement relative to the baseline scenario of~\cite{Liao2017}.}

\label{tab:SKR_Enhancement}
\begin{tabular}{|c|c|c|c|}
\hline
\boldmath$\theta$ [deg] & \makecell{\textbf{RIS} \\ \textbf{Configuration}} & \makecell{\textbf{SKR} \\ \textbf{[bits/s]}} & \makecell{\textbf{Improvement} \\ \textbf{over Baseline}} \\
\hline
\multirow{4}{*}{20} 
  & \textbf{Baseline}   & \textbf{1\,100} & -- \\
  & N = 128             & 1\,382          & +25\,\% \\
  & N = 265             & 1\,683          & +53\,\% \\
  & N = 512             & 2\,226          & +102\,\% \\
\hline
\multirow{4}{*}{80} 
  & \textbf{Baseline}   & \textbf{3\,500} & -- \\
  & N = 128             & 4\,396          & +26\,\% \\
  & N = 265             & 5\,355          & +53\,\% \\
  & N = 512             & 7\,084          & +102\,\% \\
\hline
\end{tabular}
\end{table}

\selectlanguage{english}%

\begin{figure}
\centering{}\includegraphics[width=3.5in]{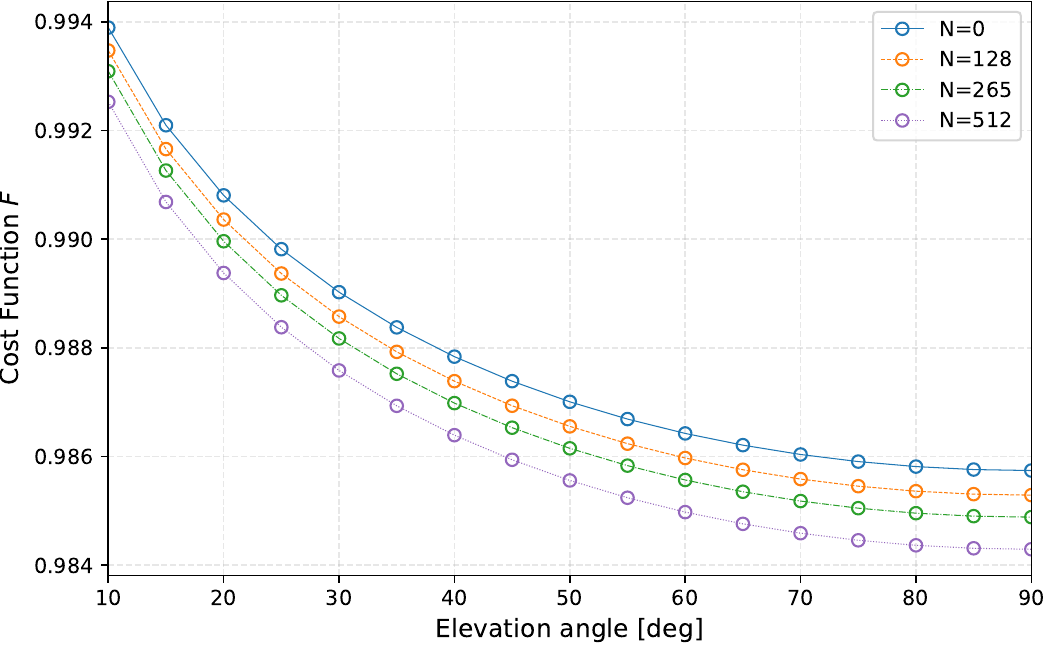}
\caption{combined QBER and SNR cost metric as a function of elevation angle\label{fig:8} }
\end{figure}

Integrating quantum and classical channel performance metrics, Fig. \ref{fig:8} presents the theoretical cost function \eqref{eq: F(f)} as a function of the satellite elevation angle for the baseline ($N\,=\,0$) and for RIS-assisted links with $N\,=\,128,265,$ and 512 elements. This composite metric, derived in Eq. \eqref{eq: F(x)}, quantifies the joint performance of the quantum and classical channels by penalizing QBER and rewarding classical S-band SNR within a balanced and  normalized framework. As the elevation angle increases, the cost function decreases monotonically, reflecting the physical improvements in both QBER and SNR due to reduced atmospheric path loss and enhanced channel conditions at higher elevations. Incorporating a RIS yields a marked reduction in the cost function across all elevation angles. Larger RIS arrays deliver greater performance gains; however, the improvements become less pronounced as the array size increases, consistent with theoretical predictions of diminishing returns imposed by phase quantization, mutual coupling, and finite aperture constraints.

The empirical trends in the figure closely follow the theoretical model, confirming that a dual-band RIS can significantly enhance the performance of integrated quantum-classical links, especially at low elevation angles where the link budget is most challenged. This validates the utility of the proposed cost function as a unified metric for optimizing future satellite communication systems that must simultaneously satisfy both
quantum security and classical reliability. 
\section{Conclusion\label{sec:7}}

This paper presents a comprehensive dual-band RIS-assisted satellite-to-ground communication system that simultaneously supports QKD at $\mathrm{850\:nm}$ and classical RF communication in the S-band. Through a comprehensive theoretical analysis incorporating realistic impairments such as atmospheric turbulence, free-space losses, pointing errors, and ionospheric effects, we developed an adaptive optimization strategy formulated as a QUBO. Simulation results provided substantial evidence of significant improvements over the performance benchmarks set by existing QKD satellite systems, notably the Micius project. Key outcomes include a reduction in QBER to approximately $\mathrm{0.7\%}$ at zenith elevation, an increase in SKR exceeding 30,000 bits/s with a 512-element RIS array, and enhancements in classical RF SNR of approximately 3 dB across all satellite elevation angles. The adaptive optimization capability demonstrated real-time robustness against environmental fluctuations, underscoring the practical applicability of the RIS system. 

The adaptive optimization framework presented here offers substantial practical promise for real-time deployment by efficiently adapting to dynamic environmental conditions. These outcomes provide a foundational advancement toward robust hybrid quantum-classical satellite communication systems, with potential for future research focused on higher-resolution RIS configurations and improved adaptive optimization methodologies.

\bibliographystyle{IEEEtran}
\bibliography{IEEEr}

\end{document}